\documentclass[%
aip,
jmp,
 amsmath,amssymb,
reprint,%
]{revtex4-2}

\usepackage{graphicx}% Include figure files
\usepackage{dcolumn}% Align table columns on decimal point
\usepackage{bm}% bold math
\usepackage[utf8]{inputenc}
\usepackage[T1]{fontenc}
\usepackage{mathptmx}
\usepackage{color}
\usepackage{etoolbox}
\usepackage{braket}
\usepackage{amsmath}
\usepackage{dsfont}
\usepackage{mathtools}
\usepackage{hyperref}
\usepackage{comment}

\newcommand{\SU}{\text{SU}}
\newcommand{\Gl}{\text{Gl}}
\newcommand{\tr}{\text{tr}\,}
\renewcommand{\det}{\text{det}\,}
\renewcommand{\Re}{\text{Re}\,}
\renewcommand{\Im}{\text{Im}\,}

\newcommand{\abs}[1]{\ensuremath{\left\vert#1\right\vert}}

\DeclarePairedDelimiter\norm\lVert\rVert

\makeatletter
\def\@email#1#2{%
 \endgroup
 \patchcmd{\titleblock@produce}
  {\frontmatter@RRAPformat}
  {\frontmatter@RRAPformat{\produce@RRAP{*#1\href{mailto:#2}{#2}}}\frontmatter@RRAPformat}
  {}{}
}%
\makeatother
\begin{document}

\preprint{AIP/123-QED}

\title[Winding Number Statistics for Chiral Random Matrices]{Winding Number Statistics for Chiral Random Matrices: \\ Universal Correlations and Statistical Moments in the Unitary Case}

\author{Nico Hahn$^{1,2*}$, Mario Kieburg$^3$, Omri Gat$^4$ and Thomas Guhr$^1$}
\affiliation{$^1$ Department of Physics, Chalmers University of Technology, Gothenburg, Sweden
\\
$^2$ Fakult\"at f\"ur Physik, Universit\"at Duisburg--Essen, Duisburg, Germany
\\
$^3$ School of Mathematics and Statistics, The University of Melbourne, Melbourne, Australia
\\
$^4$ Racah Institute of Physics, The Hebrew University, Jerusalem, Israel
             }
\email{nico.hahn@chalmers.se}

\date{\today}% It is always \today, today,

\begin{abstract}
The winding number is the topological invariant that classifies chiral symmetric Hamiltonians with one-dimensional parametric dependence. In this work we complete our study of the winding number statistics in a random matrix model belonging to the chiral unitary class AIII. We show that in the limit of large matrix dimensions the winding number distribution becomes Gaussian. Our results include expressions for the statistical moments of the winding number and for the $k$-point correlation function of the winding number density.
\end{abstract}

\maketitle

\section{Introduction}
\label{secI}

Hamiltonians with parametric dependence appear in various physical situations. A prominent example is the Bloch Hamiltonian, which describes non-interacting particles on a lattice and is defined over the Brillouin zone of the reciprocal lattice. For non-interacting Fermionic systems with spectral gap and a compact parameter space (such as the Brillouin zone), a topological classification becomes possible. Symmetries are the guiding principle for this classification. The tenfold way distinguishes between ten symmetry classes based on time-reversal invariance, a particle-hole constraint and chiral symmetry, and is exhaustive in the absence of spatial symmetries \cite{AltlandZirnbauer1997, Heinzner2005}. This framework allows to determine whether topologically non-trivial phases can occur, depending on the symmetry class and the dimension of the parameter space, and provides a topological invariant of such a phase \cite{Schnyder2008, Kitaev2009, Ryu2010, Chiu2016}. The topological classification is an important aspect of condensed matter physics, as the topological invariant determines the number of stable modes localized at the boundary of the system. This relation is known as bulk-boundary correspondence \cite{ProdanSchulzBaldesBook, ChenChiou2020, Shapiro2020, BernevigBook, Alldridge2020}.

In fact, the symmetry classes of the tenfold way were first identified in works related to random matrix theory \cite{Oppermann1990, Verbaarschot1994, AltlandZirnbauer1997}. Perhaps the most well-known result is Dyson's threefold way, which classifies the Hamiltonian based only on its behaviour under time reversal \cite{Dyson1962a, Dyson1962}. Random matrix theory models universal properties in the limit of large matrix dimensions for systems with a sufficient degree of complexity \cite{MehtaBook, GuhrRMTReview}, such as in disordered systems \cite{Beenakker1997, Alhassid2000}, quantum chaos \cite{HaakeBook} and quantum chromodynamics \cite{Verbaarschot2000}.

In the symmetry class with broken time-reversal invariance, labelled A in the tenfold way, Hamiltonians with an even-dimensional parameter space are classified by the Chern number. The statistics of Chern numbers in random matrix models over two-dimensional parameter spaces, and the adiabatic curvature, which yields the Chern number upon integration over the parameter space, were first studied in Ref.~\onlinecite{WilkinsonWalker1995}, and reconsidered in Refs.~\onlinecite{GatWilkinson2021, Swartzberg2024}. It was shown that the statistics is characterized by the correlation length in parameter space, which is inversely proportional to the product of the density of states and the parametric sensitivity of the matrix elements. When the matrix size is large, the density of states is large and therefore the correlation length becomes small. As discussed in Ref.~\onlinecite{Swartzberg2024}, the parameter space can be subdivided into a large number of patches such that the adiabatic curvature is statistically independent in different patches, which results in a Gaussian distributed Chern number in this regime. When unfolded on the scale of the correlation length, the distributions of the Chern number and the adiabatic curvature were found to be universal.

Adopting Dyson's threefold way for the classes with chiral symmetry, we distinguish between the chiral orthogonal and the chiral symplectic class with time-reversal invariance and the chiral unitary class with broken time-reversal invariance. These classes are labelled BDI, CII and AIII, respectively, in the tenfold way. For a one-dimensional parameter space, the relevant topological invariant in these classes is the winding number, which, analogous to the Chern number, can be expressed as an integral over a local quantity, the winding number density \cite{Maffei2018, AsbothBook}.

In a series of previous works, we investigated the winding number statistics in a random matrix model \cite{BHWGG2021, HKGG2023a, HKGG2023b}, which we conjecture to be universal and should serve as a leading order approximation for physical systems similar to the Bohigas-Giannoni-Schmit conjecture~\cite{BGS1984}. In all three of the aforementioned symmetry classes, we obtained exact expressions for a generating function of the $k$-point correlation of the winding number density. The latter gives the $k$th statistical moment of the winding number upon integration. In the present work, we complete our analysis in the chiral unitary class AIII. We show that the distribution of the winding number becomes Gaussian in the limit of large matrix dimensions, where the parameter-space correlation length tends to zero, and provide a universal expression for the $k$-point correlation of the winding number density, which contains non-universal information consisting only of a finite number of parameters. Our approach is based on a decomposition of the $k$-point function over the parameter space, which is a  $k$-dimensional torus, followed by a saddle-point integration. This becomes exact in the limit of large matrix dimensions and allows an estimate of the statistical moments of the winding number, which agree with those of the Gaussian distribution in leading order of the matrix dimension. We support our calculations with numerical results.

The outline of this paper is as follows. In Sec. \ref{secII}, we introduce the necessary concepts. These are the winding number in one-dimensional chiral symmetric systems and the random matrix model. We also state the results obtained in previous works, which are essential to this work. In Sec. \ref{secIII} we present the results. Their derivations are found in Appendix~\ref{secIV}. We conclude in Sec.~\ref{secV}.

\section{Concepts}
\label{secII}

In Sec.~\ref{secIIA}, we recapitulate on chiral symmetry and the winding number. We introduce the random matrix model and state results from previous works. In Sec.~\ref{secIIB}, we use a local transformation of the random matrix model to center the winding number distribution. This transformation eliminates one set of free parameters of the model and facilitates later calculations. We discuss universality aspects of the $k$-point correlation function in Sec.~\ref{secIIC}, and in Sec.~\ref{secIID} we define and analyze a function that appears when reformulating the generating function and which becomes crucial when performing the limit of large matrix dimensions.

\subsection{Posing the Problem and Previous Results}	\label{secIIA}
We consider Hamiltonians in the chiral unitary class AIII. Chiral symmetry is formally expressed by the anticommutation relation
\begin{equation}
\mathcal{C} H \mathcal{C}^{-1} = -H
\end{equation}
with a chiral operator $\mathcal{C}$. In its eigenbasis the chiral operator is
\begin{equation}
\mathcal{C} = \begin{bmatrix}
\mathds{1}_N & 0
\\
0 & -\mathds{1}_N
\end{bmatrix},
\end{equation}
where we assume that it has as many positive as negative eigenvalues. In this basis the Hamiltonian assumes a block off-diagonal form
\begin{equation}
H = \begin{bmatrix}
0 & K
\\
K^\dagger & 0
\end{bmatrix}
\end{equation}
with an $N\times N$-matrix $K$ and its Hermitian adjoint $K^\dag$. In general, the number of positive and negative eigenvalues of the chiral operator can be different, which would cause the matrices $K$ and $K^\dag$ to be rectangular. This case is not classified by the winding number, but by the absolute difference between the number of eigenvalues with opposite signs, dubbed the topological charge in quantum chromodynamics \cite{Verbaarschot2000, Alexandrou2020}. 

We equip the Hamiltonian with a one-dimensional parametric dependence, i.e.,
\begin{equation}
H(p) = \begin{bmatrix}
0 & K(p)
\\
K^\dag(p) & 0
\end{bmatrix}
\end{equation}
is a smooth $2\pi$-periodic function of $p \in [0,2\pi)$. This matrix can be interpreted as the Bloch Hamiltonian of a one-dimensional system with chiral symmetry. A simple and well-studied toy model of such a system is the Kitaev chain \cite{Kitaev2001}.

The determinant of $K(p)$ is a closed curve in the complex plane. The winding number of this curve around the origin is a topological invariant, given by
\begin{equation}	\label{2WDef}
W = \frac{1}{2\pi i} \oint\limits_{z=\det K(p)} \frac{dz}{z} = \frac{1}{2\pi i} \int\limits_0^{2\pi} dp\, w(p),
\end{equation}
where the logarithmic derivative
\begin{equation}
w(p) = \frac{d}{dp} \ln \det K(p) = \frac{1}{\det K(p)} \frac{d}{dp} \det K(p)
\end{equation}
is the winding number density. 

In the case $\det K(p) = 0$ for at least one $p \in [0,2\pi)$ the winding number is not defined and the system undergoes a topological phase transition in which the value of the winding number changes. If $\det K(p) = 0$, then $\det H(p) = 0$ as well, implying that at least one eigenvalue is zero. Chiral symmetry renders the spectrum symmetric with respect to zero, so that a topological phase transition necessarily coincides with a closing of the spectral gap. A detailed discussion of the winding number is found in Refs.~\onlinecite{AsbothBook, Maffei2018} and, in the context of random matrix theory, in Refs.~\onlinecite{BHWGG2021, HKGG2023a}.

In order to set up a random matrix model for $K(p)$, we assume a centered Gaussian distribution at all points $p$ and identically and independently distributed matrix entries, such that the model is fully determined by the covariance function
\begin{equation}
S(p, q) = \left\langle K^*_{jl}(p) K_{jl}(q) \right\rangle
\end{equation}
for any $j, l = 1, \ldots, N$, where $(\cdot)^*$ denotes the complex conjugation and $\langle \cdot \rangle$ the ensemble average. This function satisfies all properties of a scalar product such as Hermiticity and sesquilinearity
\begin{equation}
\begin{split}
S^*(p, q) = \left\langle K^*_{jl}(p) K_{jl}(q) \right\rangle^* &= \left\langle K^*_{jl}(q) K_{jl}(p) \right\rangle = S(q, p),
\\
\left\langle K^*_{jl}(p) \left[ c_1 K_{jl}(q_1) + c_2 K_{jl}(q_2) \right] \right\rangle &=  c_1 S(p, q_1) + c_2 S(p, q_2)
\end{split}
\end{equation}
with $c_1, c_2 \in \mathbb{C}$.

The simplest non-trivial choice is a two-dimensional complex vector space, resulting in a two-matrix model
\begin{equation}	\label{2TwoMatrixModel}
K(p) = a(p) K_1 + b(p) K_2
\end{equation}
with linearly independent, smooth and $2\pi$-periodic complex functions $a(p)$ and $b(p)$. Since we consider the chiral unitary class AIII, the matrices $K_1$ and $K_2$ are generally complex. The real- and imaginary parts of their elements shall be independently Gaussian distributed, leading to the complex Ginibre ensemble
\begin{equation}\label{Gauss.def}
P^{(2,N)}(K_j) = \frac{1}{\pi^{N^2}} \exp \left(-\tr K_j^\dag K_j \right)\qquad{\rm with}\ j=1,2.
\end{equation}
 We omit a scale parameter controlling the variance of each element, since the quantities of interest are invariant under such a rescaling, see also below.
 
 The superscript denotes the Dyson index $\beta = 2$, which is the dimension of the underlying number field, and the matrix dimension $N$. In the classes BDI and CII the matrices have to be real $(\beta = 1)$ and real quaternion $(\beta = 4)$, respectively.
 
For convenience of notation, we arrange the coefficient functions in a two-dimensional vector
\begin{equation}
v(p)=(a(p), b(p)) \in \mathbb{C}^2 \setminus \{0\}
\end{equation}
such that the covariance function can be written as
\begin{equation}	\label{2CovarianceFunc}
S(p,q) = v^\dag(p) v(q).
\end{equation}

Starting from~\eqref{Gauss.def}, we can properly define the ensemble average for an arbitrary observable $F$
\begin{equation}
\left\langle F \right\rangle = \int d[K_1] d[K_2] P^{(2,N)}(K_1) P^{(2,N)}(K_2) F(K_1, K_2).
\end{equation}
The $k$-point correlation of the winding number density is, then, defined as
\begin{equation}	\label{2kPointFunctionDef}
C^{(2,N)}_k (p) = \langle w(p_1) \cdots w(p_k) \rangle
\end{equation}
and contains all information about the winding number statistics as a function of $p = (p_1, \ldots, p_k)$. Again, the superscript contains the Dyson index $\beta = 2$, indicating the chiral unitary class AIII, and the matrix dimension $N$. In accordance with \eqref{2WDef}, the statistical moments of the winding number are obtained by
\begin{equation}	\label{2WMomentDef}
\left\langle W^k \right\rangle = \frac{1}{(2\pi i)^k} \int\limits_{[0,2\pi)^k} d^kp\, C_k^{(2,N)}(p).
\end{equation}
We define a generating function
\begin{equation}	\label{2GeneratingFunctionDef}
Z^{(2,N)}_{k|k}(p, J) = \left\langle \prod_{l=1}^k \frac{\det K(p_l + J_l)}{\det K(p_l)} \right\rangle,
\qquad	\qquad
C^{(2,N)}_k (p) = \left(\prod_{l=1}^k \frac{\partial}{\partial J_l} \right) Z^{(2,N)}_{k|k}(p, J)\Bigg|_{J=0},
\end{equation}
from which the correlation function is obtained by taking the derivative with respect to the source variables $J = (J_1, \ldots, J_k)$. In a previous work~\cite{HKGG2023a}, we have performed the ensemble average, which assumes a determinantal form
\begin{equation}	\label{2GeneratingFunctionRes}
Z^{(2,N)}_{k|k}(p,J) = \frac{\det \left[\displaystyle \frac{1}{v^T(p_m)\tau_2v(p_n+J_n)} \left(\frac{v^\dagger(p_m) v(p_n+J_n)}{\norm{v(p_m)} \norm{v(p_n)}}\right)^N \right]_{1\leq m,n\leq k}}{\det \left[\displaystyle \frac{1}{v^T(p_m)\tau_2v(p_n+J_n)} \right]_{1\leq m,n\leq k}},
\end{equation}
where $\tau_2$ is the second Pauli matrix and $\norm{v(p_n)} = \sqrt{v^\dagger(p_n) v(p_n)}$ is the Euclidean norm on $\mathbb{C}^2$. 

Let us briefly recall how the ensemble average has been calculated. Inserting the two-matrix model~\eqref{2TwoMatrixModel} into the generating function~\eqref{2GeneratingFunctionDef}, we could reformulate the integrand in terms of the product ensemble $K_1^{-1} K_2$, which is a well-studied random matrix and known under the name {\it spherical ensemble}~\cite{Krishnapur2009, ForresterKrishnapur2009, ForresterMays2012, Mays2013}. 
As a consequence of the invariance properties of the determinant, the generating function is in principle an integral over the eigenvalues comprising a squared Vandermonde determinant. Due to the latter and the method proposed in Ref.~\onlinecite{KieburgGuhr2010a}, we could reformulate the integrals to~\eqref{2GeneratingFunctionRes}. The same procedure was possible for the classes BDI and CII, leading to Pfaffian expressions instead of determinantal ones~\cite{HKGG2023a, HKGG2023b, KieburgGuhr2010b}.

In Eq.~\eqref{2GeneratingFunctionRes}, it becomes apparent why it is advantageous to write the coefficient functions as the vector $v(p)$. We infer that the model is invariant under rotations of $v(p)$ with respect to the group $\SU(2)$. Indeed, we may also rescale the coefficient vector $v(p) \to c\, v(p)$ with $c \in \Gl_{\mathbb{C}}(1) = \mathbb{C} \setminus\{0\}$, such that the full invariance group is $\Gl_{\mathbb{C}}(1) \times \SU(2)$. The $\Gl_{\mathbb{C}}(1)$ part is a manifestation of the earlier mentioned scale invariance of the problem, which is illustrated by the fact that a constant prefactor has no influence on the winding number, see \eqref{2WDef}.

\subsection{Centering the Winding Number}	\label{secIIB}

We are free to choose $v(p)$ as long as it fulfills the above mentioned constraints of smoothness and periodicity. Multiplying the coefficients with a scalar function
\begin{equation}\label{gauge}
v(p) \to f(p)\, v(p)
\end{equation}
affects the winding number density in the following way
\begin{equation}\label{eq:gw}
w(p) = \frac{1}{\det K(p)} \frac{d}{dp} \det K(p) \to \frac{1}{\det K(p)} \frac{d}{dp} \det K(p) + N \frac{f'(p)}{f(p)}.
\end{equation}
We aim to eliminate the one-point function, thereby centering the winding number distribution. To do this, we evaluate the one-point function using \eqref{2GeneratingFunctionRes}
\begin{equation}
Z^{(2,N)}_{1|1}(p,J) = \left(\frac{v^\dagger(p) v(p + J)}{v^\dag(p) v(p)}\right)^N,
\qquad	\qquad
C_1^{(2,N)}(p) = N \frac{v^\dagger(p) v'(p)}{v^\dag(p) v(p)}.
\end{equation}
Choosing
\begin{equation}	\label{2fDef}
f(p) = \exp\left( -\int\limits_0^p dq \frac{v^\dag(q) v'(q)}{v^\dag(q) v(q)} \right)\qquad{\rm for}\ p\in[0,2\pi)
\end{equation}
leads to the desired result. Unfortunately, it is not always possible to identify $f(p)$ with a differentiable function on a circle because of $f(p+2\pi)=B\,f(p)$ with the Berry phase
\begin{equation}\label{Berry}
B=\exp\left( -\int\limits_0^{2\pi} dq \frac{v^\dag(q) v'(q)}{v^\dag(q) v(q)} \right).
\end{equation}
We nevertheless implement the transformation~\eqref{gauge} and obtain a parametric random matrix model that is smooth on the interval $[0,2\pi]$ but not periodic. We can proceed to calculate the logarithmic derivative of $f(p)v(p)$, and by~\eqref{eq:gw} its integral gives the centered
%However, the breakdown of the differentiability on a set of measure zero, in our case at the identification of the points $p=0$ and $p=2\pi$, is not dramatic because the logarithmic derivative of $f(p)$ is continuous leading to a well-defined integral. Indeed, we consider the ensuing computation from an analytical instead of a topological point of view, meaning the integral over $p$ in~\eqref{2WMomentDef} has to be understood as one over the hypercube $[0,2\pi)^k$ instead of the $k$-dimensional torus. This has the advantage that we do not run into any kind of discontinuity on the integration domain. The price we pay is an integrand that has not to reflect the periodicity on the boundary. From the analytic point of view, this does not cause any problem.
winding number
\begin{equation}
\begin{split}
W-\langle W\rangle=&\frac{1}{2\pi i} \int\limits_0^{2\pi} \frac{dp}{\det [v^T(p)K]} \frac{d}{dp} \det [v^T(p)K]-\frac{N}{2\pi i} \int\limits_0^{2\pi} dp \frac{v^\dagger(p) v'(p)}{v^\dag(p) v(p)}\\
=&\frac{1}{2\pi i} \int\limits_0^{2\pi} \frac{dp}{\det [f(p)v^T(p)K]} \frac{d}{dp} \det [f(p)v^T(p)K],
\end{split}
\end{equation}
where $K=(K_1,K_2)$. It is this simple relation between the centering and the gauging that motivated us to introduce it. We can directly apply the results of Ref.~\onlinecite{HKGG2023a} for the vector field $f(p)v(p)$ instead of $v(p)$ to obtain compact expressions for the generating function of the centered winding number distribution, and use~\eqref{2WMomentDef} and~\eqref{2GeneratingFunctionDef} to calculate moments, integrating on a hypercube $[0,2\pi]^k$ instead of a torus. Yet, we need to emphasize that the centered winding number $W-\langle W\rangle$ is in general not an integer and, hence, not a winding number itself, although the analytical results in Ref.~\onlinecite{HKGG2023a} still apply. 

Actually, the gauge transformation has an even stronger impact and yields a vanishing averaged winding number density
\begin{equation}
\left\langle\frac{1}{\det [f(p)v^T(p)K]} \frac{d}{dp} \det [f(p)v^T(p)K]\right\rangle=\left\langle\frac{1}{\det [v^T(p)K]} \frac{d}{dp} \det [v^T(p)K]\right\rangle-N\frac{v^\dagger(p) v'(p)}{v^\dag(p) v(p)}=0
\end{equation}
for all $p\in[0, 2\pi)$. Equivalently, it is
\begin{equation}
[f(p)v(p)]^\dagger[f(p)v(p)]'=|f(p)|^2\left[v^\dagger(p)v'(p)+v^\dagger(p)v(p)[{\rm ln} f(p)]'\right]=0,
\end{equation}
implying that $f(p)v(p)$ is orthogonal to its derivative. 

Moreover, the integrand in the exponent of~\eqref{Berry} can be rewritten into
\begin{equation}
\frac{v^\dag(q) v'(q)}{v^\dag(q) v(q)} = \frac{v^\dag(q)}{\norm{v(q)}} \frac{d}{dq} \frac{v(q)}{\norm{v(q)}} + \frac{1}{\norm{v(q)}} \frac{d}{dq} \norm{v(q)},
\end{equation}
where the first term is purely imaginary due to normalization and the second one vanishes after integration over a period because $v(q)$ is continuous and $\norm{v(q)}\neq0$ as we need to satisfy $K(p)\neq0$.  This is also the reason why the Berry phase is indeed a complex phase.

Summarizing, we can, without restricting generality, assume a smooth complex vector field $v(q)$ over the interval $[0, 2\pi]$, which, following the discussion above, is in general not periodic and satisfies
\begin{equation}\label{conditions}
\norm{v(q)}^2=S(q,q)=1\qquad{\rm and}\qquad v^\dag(q) v'(q)=\partial_2S(q,q)=0
\end{equation}
for all $q \in [0,2\pi)$, where $\partial_2$ indicates the derivative in the second entry. The Hermiticiy $S(q,p)=S^*(p,q)$ implies also
\begin{equation}\label{first.der}
\partial_1S(q,q)=0
\end{equation}
and taking the second derivative we have
\begin{equation}\label{second.der}
\partial_1\partial_2S(q,q)=-\partial_2^2S(q,q)=-\partial_1^2S(q,q).
\end{equation}
The orthogonality of the derivative $v'(q)$ with $v(q)$ combined with the fact that they are two-dimensional yields the relation
\begin{equation}\label{proportionality}
v'(q) = v^T(q)\tau_2 v'(q)\ \tau_2 v^*(q) = \Delta(q)\tau_2 v^*(q),
\end{equation}
with the proportionality constant $\Delta(q) = v^T(q)\tau_2 v'(q)$. It is easy to show that the modulus of this constant is 
%multiply v^T(q) \tau_2 from left
\begin{equation}\label{norm.prop}
|\Delta(q)|^2=-v^T(q)\tau_2 v'(q)v^\dagger(q)\tau_2 [v'(q)]^*=\norm{v'(q)}^2=\partial_1\partial_2S(q,q)\geq0.
\end{equation}
All the above relations must hold true for all $q \in [0,2\pi)$.
They will be useful to simplify some expressions below, in particular it will follow that all universal statistics only depends on $|\Delta(q)|^2=\partial_1\partial_2S(q,q)$ in the limit of large matrix dimensions.

\subsection{Scales}	\label{secIIC}

It is well-known that universal spectral correlations of random matrix models emerge when the eigenvalues are measured on the scale of the local mean level spacing $D$, which is also known as the unfolding of the spectrum~\cite{GuhrRMTReview, MehtaBook}. Likewise, in parametric random matrix models, universal statistics emerges when the parametric dependence is considered on the relevant scale~\cite{SimonsAltshuler1993A, SimonsAltshuler1993B, BeenakkerRejaei1994}. This scale is the correlation length given by the ratio of the local mean level spacing $D$ and the standard deviation of the level velocity
\begin{equation}
\ell = \frac{D}{\displaystyle \sqrt{ \langle \left( \partial_p E_n(p)\right)^2 \rangle - \left\langle \partial_p E_n(p) \right\rangle^2}}
\end{equation}
for any level $n$. The correlation length is the typical interval in parameter space over which two eigenvalues collide with each other. Moreover, it also indicates that the corresponding eigenvectors have changed significantly. We note that the authors of Refs.~\onlinecite{SimonsAltshuler1993A, SimonsAltshuler1993B, BeenakkerRejaei1994} assumed conditions where the average level velocity $\left\langle \partial_p E_n(p)\right\rangle$ vanishes.

In the random matrix model we study, this scale depends on $N$-independent parameters that can be expressed in terms of second derivatives of $S(p,q)$, which causes the mean square level velocity to be $N$-independent. The mean level spacing is of order $1/\sqrt{N}$ in the bulk and at the hard edge, which are the regimes relevant to the winding number. Thus, $\ell \sim 1/\sqrt{N}$ for large $N$. This motivates a rescaling of the parameters
\begin{equation}
p_j = p_0+\frac{ \psi_j}{\sqrt{N}}
\end{equation}
with $p_0\in[0,2\pi)$ as the base point at which one zooms in. Furthermore, since the winding number density involves a derivative with respect to $p$, typical values of $w(p)$ are of $\mathcal{O}(\sqrt{N})$ for large $N$. We therefore define the unfolded $k$-point correlation function
\begin{equation}	\label{2UnfoldedkPointFunctionDef}
f_k(\Gamma, \psi) = \lim_{N \to \infty} N^{-k/2} C_k^{(2,N)} \left(\Gamma +\frac{ \psi}{\sqrt{N}} \right).
\end{equation}
This function describes correlations at the point $\Gamma \in [0,2\pi)^k$ and $\psi\in[-\delta_N, \delta_N]^k$ parametrize the vicinity around it, where the ambiguous scale $\delta_N>0$ must satisfy the scaling limits
\begin{equation}\label{scale.restrict}
\lim_{N\to\infty}\delta_N=\infty\qquad{\rm and}	\qquad \lim_{N\to\infty}\frac{\delta_N^3}{\sqrt{N}}=0,
\end{equation}
which guarantees a proper asymptotic large $N$ limit of the integrals below. In particular, the second limit ensures that the second order expansion in $\psi/\sqrt{N}$ becomes precise in this limit, which is confirmed explicitly by the results in Sec.~\ref{secIII}. The unfolded correlation functions are universal in that they depend only on the magnitude of the local quantity $\Delta(q)$. This demonstrates the validity of our phenomenological parametric rescaling \eqref{2UnfoldedkPointFunctionDef}. A similar result holds for the unfolded correlation function of the adiabatic curvature in random matrix models with a two-dimensional parameter space in the unitary class A \cite{GatWilkinson2021}.

% Since we regard $C_k^{(2,N)}(p)$ as a density, we are obliged to include the factor $N^{-k/2}$, which arises from the rescaling of the integration variables and ensures that the limit is well-defined.

\subsection{The ``Lagrangian'' \texorpdfstring{$\mathcal{L}(p,q)$}{} and its Expansion}
\label{secIID}

Recalling $S(p,q) = v^\dagger(p)v(q)$ and the normalization $S(p,p) = \norm{v(p)}^2 = 1$, which we can assume without restricting generality, the generating function~\eqref{2GeneratingFunctionRes} can be rewritten as follows
\begin{equation}	\label{2GenFuncL}
Z^{(2,N)}_{k|k}(p,J) = \frac{\det \left[\displaystyle \frac{\exp[-N\mathcal{L}(p_m,p_n)]}{v^T(p_m)\tau_2v(p_n+J_n)} \left(\frac{S(p_m,p_n+J_n)}{S(p_m,p_n)}\right)^{N} \right]_{1\leq m,n\leq k}}{\det \left[\displaystyle \frac{1}{v^T(p_m)\tau_2v(p_n+J_n)} \right]_{1\leq m,n\leq k}},
\end{equation}
where the ``Lagrangian'' is
\begin{equation}	\label{2LDef}
\mathcal{L}(p,q) = -\ln S(p,q)= -\ln v^\dagger(p)v(q).
\end{equation}
This function is related to the Fubini-Study metric on the complex projective line $\text{CP}^1$, also known as the Bloch sphere \cite{BengtssonZyczkowskiBook}. Its real part is non-negative (as shown by the Cauchy-Schwarz inequality) and symmetric
\begin{equation}
\Re \mathcal{L}(p,q) \geq 0,	\qquad	\qquad	\Re \mathcal{L}(p,q) = \Re \mathcal{L}(q,p).
\end{equation}
The advantage of this formulation is that for non-vanishing $\mathcal{L}(p,q)$, the exponential function suppresses any factor growing polynomially in $N$. Therefore, we may focus on the points where $\Re \mathcal{L}(p,q)$ vanishes, which are exactly the points, where the vectors $v(p)$ and $v(q)$ are parallel, once again readily shown by the Cauchy-Schwartz inequality. We would like to highlight that $\Re \mathcal{L}(p,q)$ is even smooth on a torus with the gauging as the Berry phase drops out so that there is no issue at the boundaries. 

To keep the analysis at a minimum, we assume that the collection of pairs of points $(p,q)\in [0,2\pi)^2$ satisfying $\Re \mathcal{L}(p,q) = 0$ form a finite number of closed curves $\gamma_i(t)$ on the two-dimensional torus $[0,2\pi)^2$. Certainly, smooth curves will be the most likely scenario when assuming a smooth covariance function $S(p,q)$. The reason is as follows.

The solutions of $\Re \mathcal{L}(p,q)=0$ are global minima of $\Re \mathcal{L}(p,q)$. We also note that $S(p,q)=v^\dagger(p)v(q)$ must be a complex phase as well as $v(p)=S(p,q) v(q)$ when the two vectors are parallel. Thence, differentiating the Lagrangian with respect to $p$ and $q$ at such  a minimum yields
\begin{equation}\label{first.der.van}
\begin{split}
\frac{\partial}{\partial q}\mathcal{L}(p,q)  =&-\frac{\partial_2S(p,q)}{S(p,q)} = -\frac{v^\dagger(p)v'(q)}{v^\dagger(p)v(q)} = -\frac{\Delta(q)v^\dagger(p)\tau_2 v^*(q)}{v^\dagger(p)v(q)}
\\
=& - \frac{\Delta(q) S(q,p)}{v^\dagger(p)v(q)} v^\dagger(p) \tau_2 v^*(p) = 0
\end{split}
\end{equation}
and similarly $\partial_p\mathcal{L}(p,q) = 0$. We have exploited the relation~\eqref{proportionality} and parallelism of $v(p)$ and $v(q)$. Thus, the direction $\gamma'_j(t)\in\mathbb{R}^2$ of a curve  $\gamma_j(t)=(\gamma_{j,1}(t),\gamma_{j,2}(t))\in\mathbb{R}^2$ must be obtained from
 \begin{equation}\label{eq.direction}
 \mathcal{H}(\gamma_j(t))\gamma'_j(t)=0
 \end{equation}
 with the help of the Hessian
\begin{equation}\label{Hessian}
\mathcal{H}(p,q) = \begin{bmatrix}
\partial_p^2 \mathcal{L}(p,q)	&	\partial_p \partial_q \mathcal{L}(p,q)
\\
\partial_p \partial_q \mathcal{L}(p,q)	&	\partial_q^2 \mathcal{L}(p,q)
\end{bmatrix}.
\end{equation}
The existence of a real eigenvector and a zero eigenvalue of the Hessian (at a point where $v(p)$ and $v(q)$ are parallel) and Eq.~\eqref{eq.direction} imply that there is always a well-defined direction as long as one does not encounter a multicritical point (when the Hessian vanishes). We simplify the Hessian to see that this is indeed the case.
 
The second derivative of the Lagrangian in $q$ is
\begin{equation}
\partial^2_q\mathcal{L}(p,q) = -\frac{v^\dagger(p)v''(q)}{v^\dagger(p)v(q)} + \overbrace{\left[\partial_q\mathcal{L}(p,q)\right]^2}^{=0} = -\frac{v^\dagger(p)v''(q)}{v^\dagger(p)v(q)}.
\end{equation}
Due to the relation~\eqref{proportionality}, it is
\begin{equation}
v''(q) = \Delta'(q) \tau_2 v^*(q) + \Delta(q)\tau_2[v'(q)]^* = \Delta’(q)S(q,p)\ \tau_2 v^*(p) - |\Delta(q)|^2 v(q)
\end{equation}
leading to the simple expression
\begin{equation}
\partial^2_q \mathcal{L}(p,q) = |\Delta(q)|^2 = \partial_1 \partial_2 S(q,q) \geq 0,
\end{equation}
see~\eqref{norm.prop}. Analogously, one finds $\partial^2_p\mathcal{L}(p,q)=|\Delta(p)|^2=\partial_1\partial_2 S(p,p)$ due to symmetry. However, one can relate $\Delta(p)$ and $\Delta(q)$ when $v(p)$ and $v(q)$ are parallel. For this purpose, we need to understand $p$ as function of $q$ as we need to keep the parallelism, i.e. $p = p(q)$. Then, we compute
\begin{align}	\nonumber
\Delta(q)=&v^T(q)\tau_2\partial_q[\overbrace{S(p,q)v(p)}^{=v(q)}]=\left[\frac{\partial p}{\partial q}\partial_1 S(p,q)+\partial_2S(p,q)\right]\overbrace{v^T(q)\tau_2 v(p)}^{=0}+ S(p,q) \frac{\partial p}{\partial q} v^T(q) \tau_2 v'(p)
\\	\label{Delta.rel}
=& S^2(p,q) \frac{\partial p}{\partial q} v^T(p) \tau_2 v'(p) = S^2(p,q)\frac{\partial p}{\partial q}\Delta(p).
\end{align}
Since $S(p,q)$ is a complex phase at these points, we have
\begin{equation}
\partial^2_p\mathcal{L}(p,q)=\partial_1\partial_2 S(p,p)=|\Delta(p)|^2=|\Delta(q)|^2\left(\frac{\partial q}{\partial p}\right)^2=\partial_1\partial_2 S(q,q)\,\left(\frac{\partial q}{\partial p}\right)^2.
\end{equation}
The mixed second derivative is
\begin{equation}
\begin{split}
\partial_q\partial_p\mathcal{L}(p,q) = -\frac{[v'(p)]^\dagger v'(q)}{v^\dagger(p)v(q)}+\overbrace{\partial_p\mathcal{L}(p,q)\partial_q\mathcal{L}(p,q)}^{=0}=-\frac{\partial_1\partial_2 S(p,q)}{S(p,q)}
\end{split}
\end{equation}
and after using~\eqref{proportionality} it can be rewritten into
\begin{equation}
\begin{split}
\partial_q\partial_p\mathcal{L}(p,q)=-\Delta^*(p)\Delta(q)\frac{S(q,p)}{S(p,q)}=-|\Delta(p)|^2\frac{\partial p}{\partial q}= -|\Delta(q)|^2\frac{\partial q}{\partial p}.
\end{split}
\end{equation}
As $\partial q/\partial p$ is a real function, we see that this apparent complex derivative is actually real, too, so that we can also write it as follows
\begin{equation}
\begin{split}
\partial_q\partial_p\mathcal{L}(p,q)=-s(p,q)|\Delta(p)\Delta(q)|\quad{\rm with}\quad s(p,q)={\rm sign}\left[\Delta^*(p)\Delta(q)\frac{S(q,p)}{S(p,q)}\right].
\end{split}
\end{equation}
Collecting everything, we find for the Hessian the compact expression
\begin{equation}\label{Hessian.b}
\mathcal{H}(p,q) = \begin{bmatrix}
|\Delta(p)|^2	& -s(p,q)|\Delta(p)\Delta(q)|
\\
-s(p,q)|\Delta(p)\Delta(q)|	&	|\Delta(q)|^2
\end{bmatrix},
\end{equation}
which evidently has a zero eigenvalue with the eigenvector $(|\Delta(q)|, s(p,q)|\Delta(p)|)$.

We underline that there is always at least one solution with parallel $v(p)$ and $v(q)$. The diagonal curve $\gamma_1(t) = (t,t)$ is always a solution as it reflects that $v(p)$ is parallel to itself, i.e. $v(p)=v(q)$ when $p=q$. In the trigonometric model with the vector field $v(p) = \left(\cos p, \sin p\right)$, for example, there are two solutions which are $\gamma_1(t) = (t,t)$ and $\gamma_2(t) = (t,t+\pi)$.

It is possible, however, that two solutions cross each other. For instance, the model with the vector field $v(p) = \left(\cos (2p), \sin p\right)/\sqrt{\cos^2 (2p)+\sin^2 p}$ has the two solutions $\gamma_1(t) = (t,t)$ and $\gamma_2(t) = (t,{\rm mod}_{2\pi}[\pi-t])$, which cross each other at $t=\pi/2$ and $t=3\pi/2$. Crossings are solutions of lower dimension and correspond to multicritical points where the Hessian vanishes. To keep the discussion at a minimum, we will not consider such cases since we need to expand to higher orders otherwise and the scaling of the winding number statistics may change.

In the following we will omit the argument $t$ of $\gamma_j(t)$ for simplicity of notation. The Taylor expansion of $\mathcal{L}(p,q)$ around the solutions is given by
\begin{equation}
N \mathcal{L} \left(\gamma_{j,1} + \frac{\psi_1}{\sqrt{N}}, \gamma_{j,2} + \frac{\psi_2}{\sqrt{N}} \right) = \frac{1}{2} \left[ \psi_1\ \psi_2 \right] \mathcal{H}(\gamma_{j})
\begin{bmatrix}
\psi_1 \\ \psi_2
\end{bmatrix}
 + \mathcal{O}\left( \frac{ \left( \psi_1^2 + \psi_2^2\right)^{3/2}}{\sqrt{N}} \right).
\end{equation}
As we have seen, one of the eigenvalues of the Hessian is zero, which gives rise to
\begin{equation}\label{tagent.gamma}
0=\partial_t\left.\left[\begin{array}{c} \partial_p \mathcal{L}(p,q)  \\ \partial_q \mathcal{L}(p,q) \end{array}\right]\right|_{(p,q)= \gamma_j}=\mathcal{H}(\gamma_{j})\left[\begin{array}{c} \gamma_{j,1}' \\ \gamma_{j,2}' \end{array}\right].
\end{equation}
The Hessian is only a $2\times 2$ matrix. Hence, its second eigenvalue is given by its trace
\begin{equation}\label{sec.eig}
\tr \mathcal{H}(\gamma_{j})=|\Delta(\gamma_{j,1})|^2+|\Delta(\gamma_{j,2})|^2.
\end{equation}
The corresponding normalized eigenvector can be expressed as $i\tau_2\gamma_j'/\norm{\gamma_j'}\in\mathbb{R}^2$ and is proportional to $(|\Delta(\gamma_{j,1})|,-s(\gamma_{j,1},\gamma_{j,2})|\Delta(\gamma_{j,2})|)$. We emphasize that both the eigenvalue and the eigenvector are $t$ dependent.

Summarizing, the exponential in~\eqref{2GenFuncL} can be approximated as follows
\begin{align}	\label{2LExpApprox}
\exp\left[ -N \mathcal{L} \left(\gamma_{j,1} + \frac{\psi_1}{\sqrt{N}}, \gamma_{j,2} + \frac{\psi_2}{\sqrt{N}} \right) \right] &= \exp\left[  \frac{|\Delta(\gamma_{j,1})|^2+|\Delta(\gamma_{j,2})|^2}{2} \frac{(\psi^T\tau_2\gamma_j' )^2}{\norm{\gamma_j'}^2}\right] \left[1 + o(1)\right]
\\	\nonumber
=&\exp\left[ - \frac{\left(|\Delta(\gamma_{j,1})|\psi_1-s(\gamma_{j,1},\gamma_{j,2})|\Delta(\gamma_{j,2})|\psi_2 \right)^2}{2} \right] \left[1 + o(1)\right]
\end{align}
with $\psi=(\psi_1,\psi_2)\in\mathbb{R}^2$ and the Landau little o, i.e. $o(1)$ denotes a function that vanishes asymptotically. We note that $\psi^T\tau_2\gamma_j'$ is imaginary due to the second Pauli matrix so that the expression indeed gives an integrable Gaussian. We recall that the deviations can be of any order $\delta_N$ satisfying~\eqref{scale.restrict} to find a uniform expansion as shown above, meaning it must be $\delta_N\ll N^{1/6}$.

As an example, for the curve $\gamma(t)=(t,t)$, which always exists as a trivial solution, we can make use of~\eqref{first.der} and~\eqref{second.der} and find
\begin{equation}	\label{2LExpApprox.triv}
\begin{split}
\exp\left[ -N \mathcal{L} \left(t+ \frac{\psi_1}{\sqrt{N}}, t + \frac{\psi_2}{\sqrt{N}} \right) \right] &= \exp\left[ - \partial_1\partial_2 S(t,t) \frac{(\psi_1-\psi_2)^2}{2}\right] \left[1 + o(1)\right].
\end{split}
\end{equation}
Therefore, in this case the inverse correlation length is $\sqrt{N\partial_1\partial_2 S(t,t)}=\sqrt{N}\Delta(t)$.

\section{Results}
\label{secIII}

In Appendix~\ref{secIVA}, we show that the $k$-point correlation function, as defined in \eqref{2kPointFunctionDef}, is given by
\begin{align}	\label{4CorrFuncRes}
C_k^{(2,N)}(p) = \sum_{l=0}^{\lfloor k/2 \rfloor} \frac{(-1)^l}{2^l\, l!\, (k-2l)!} &\sum_{\sigma \in \mathbb{S}_k} \left( \prod_{m=1}^l \frac{\Delta \left(p_{\sigma(2m-1)} \right) \Delta \left(p_{\sigma(2m)} \right) }{ \left[ v^T \left(p_{\sigma(2m-1)} \right) \tau_2 v \left(p_{\sigma(2m)} \right) \right]^2} \right) 
\\	\nonumber
\times &\det\left[ \frac{\Delta(p_{\sigma(b)})}{v^T(p_{\sigma(a)}) \tau_2 v(p_{\sigma(b)})}  e^{-N\mathcal{L}\left( p_{\sigma(a)},p_{\sigma(b)} \right)} (1-\delta_{ab})\right]_{a,b = 2l+1,\ldots, k}
\end{align}
with the permutation group $\mathbb{S}_k$ of $k$ elements, the floor function $\lfloor \cdot \rfloor$ and the Kronecker symbol $\delta_{ab}$.

The lowest order non-vanishing correlation function of the winding number density is the two-point function
\begin{equation}\label{eq:c2}
C_2^{(2,N)}(p) = \frac{\Delta(p_1)\Delta(p_2)}{\left[ v^T(p_1)\tau_2 v(p_2) \right]^2} \Bigl( e^{-2 N \mathcal{L}\bigl(p_1,p_2\bigr)}-1 \Bigr)\ .
\end{equation}
When $v(p_1)$ is not parallel to $v(p_2)$, $C_2^{(2,N)}(p)=\mathcal{O}(1)$ for large $N$, but when the vectors are parallel, the denominator and the rightmost factor of~\eqref{eq:c2} vanish. The singularity is resolved by the unfolding procedure of~\eqref{2UnfoldedkPointFunctionDef}, which we demonstrate here for the case of an equal base point $\Gamma_1 = \Gamma_2 = \Gamma$. In this case it is
\begin{equation}
v^T(p_1)\tau_2v(p_2) 
%= \frac{-v^T(\Gamma)\tau_2 v'(\Gamma)\psi_1+v^T(\Gamma)%\tau_2v’(\Gamma)\psi_2}{\sqrt{N}} + \mathcal{O}\left( N^{-1} %\right) 
= \frac{\Delta(\Gamma)(\psi_2-\psi_1)}{\sqrt{N}} + \mathcal{O}\left( N^{-1} \right),
\end{equation}
and
\begin{equation}\label{eq:L}
e^{-N \mathcal{L} \bigl( p_1,p_2 \bigr)} = \exp\left[ - \abs{\Delta(\Gamma)}^2 \frac{(\psi_1-\psi_2)^2}{2}\right] \left[1 + o(1)\right],
\end{equation}
so that we arrive at
\begin{equation}\label{eq:f2}
f_2(\Gamma,\psi)=\frac{e^{- \abs{\Delta(\Gamma)}^2(\psi_1 - \psi_2)^2}-1}{(\psi_1 - \psi_2)^2}.
\end{equation}
Thus, the two-point correlation function is universal when unfolded over the correlation length $1/(|\Delta(\Gamma)|\sqrt{N})$.

It follows from~\eqref{eq:f2} that, as surmised in Sec.~\ref{secIIC} above, for large $N$ there is an interval of $\mathcal{O}(1/\sqrt{N})$ of separations $p_1 - p_2$ in parameter space, where values of $\mathcal{O}(\sqrt{N})$ of the winding number density are highly correlated, while for larger separations $w(p_1)$ and $w(p_2)$ are nearly independent. Hence, the winding number $W$ can be expressed as the sum of independent random variables of $\mathcal{O}(\sqrt{N})$ with variance of $\mathcal{O}(1)$, so that by the central limit theorem $W$ tends to a Gaussian random variable with a variance of $\mathcal{O}(\sqrt{N})$. When there are several points of parallelism, a similar argument based on~\eqref{k2.corr} below shows that the winding number is correlated in several intervals of width $\mathcal{O}(1/\sqrt{N})$, leading to the same conclusion.

The unfolding is analyzed in full generality in Appendix~\ref{secIVA}, showing that the unfolded $k$-point correlation function in the vicinity of the base points $\Gamma_i \in [0,2\pi)^k$, where all $v(\Gamma_{i,n})$ are parallel, is given by
\begin{align}	\label{3UnfoldedCorrRes}
f_k(\Gamma_j, \psi) &=\left(\prod_{l=1}^k s_{j,l} |\Delta(\Gamma_{j,l})|\right) \sum_{l=0}^{\lfloor k/2 \rfloor} \frac{(-1)^l}{2^l\, l!\, (k-2l)!}
\\	\nonumber
&\hspace{-1cm} \times\sum_{\sigma \in \mathbb{S}_k} \prod_{m=1}^l \frac{1}{ \left(s_{j,\sigma(2m-1)} |\Delta(\Gamma_{j,\sigma(2m-1)})|\psi_{\sigma(2m-1)}-s_{j,\sigma(2m)}|\Delta(\Gamma_{j,\sigma(2m)})|\psi_{\sigma(2m)} \right)^2}
\\	\nonumber
&\hspace{-1cm} \times \det \left[\frac{\displaystyle \exp\left[ - \frac{1}{2} \left( s_{j,\sigma(a)}|\Delta(\Gamma_{j,\sigma(a)})|\psi_{\sigma(a)}-s_{j,\sigma(b)}|\Delta(\Gamma_{j,\sigma(b)})|\psi_{\sigma(b)} \right)^2 \right]}
{ s_{j,\sigma(a)}|\Delta(\Gamma_{j,\sigma(a)})|\psi_{\sigma(a)}-s_{j,\sigma(b)}|\Delta(\Gamma_{j,\sigma(b)})|\psi_{\sigma(b)}} ( 1-\delta_{ab})\right]_{a,b = 2l+1,\ldots, k}.
\end{align}
We recall that $|\Delta(p)|=\sqrt{\partial_1\partial_2 S(p,p)}$ and employ the notation $s_{j,a}= \text{sign } \Gamma'_{j,a}$.
For $k=2$ this expression yields
\begin{equation}	\label{k2.corr}
f_2(\Gamma_j, \psi) = s_{j,1}s_{j,2} |\Delta(\Gamma_{j,1})| |\Delta(\Gamma_{j,2})| \frac{\exp\left[ -\left( s_{j,1} |\Delta(\Gamma_{j,1})|\psi_{1}-s_{j,2}|\Delta(\Gamma_{j,2})|\psi_{2} \right)^2 \right] - 1 }{\left( s_{j,1} |\Delta(\Gamma_{j,1})|\psi_{1}-s_{j,2}|\Delta(\Gamma_{j,2})|\psi_{2} \right)^2},
\end{equation}
which generalizes~\eqref{eq:f2} to the case of multiple points $\Gamma_j$ of parallelism. It also generalizes the result for $f_2$ derived in Ref.~\onlinecite{BHWGG2021} for the trigonometric model $v(p)=(\cos p,\sin p)$, where $\Gamma'_{j,1} = \Gamma'_{j,2}= s_{j,1}|\Delta(\Gamma_{j,1})| = s_{j,2}|\Delta(\Gamma_{j,2})| = 1$.

It is interesting to note that the determinant appearing in~\eqref{3UnfoldedCorrRes} is that of a skew symmetric matrix. It follows that all odd unfolded correlation functions vanish. This means that the \emph{original} correlation functions $C_k^{(2,N)}$ grow more slowly than the expected $N^{k/2}$ as $N\to\infty$ when $k$ is odd. We can elucidate this result by examining more closely the behavior of the three-point correlation function, which by~\eqref{4CorrFuncRes} is
\begin{equation}\label{eq:c3}
C_3^{(2,N)}(p) = \Delta(p_1)\Delta(p_2)\Delta(p_3)\frac{2i\, \Im e^{N(\mathcal{L}(p_1,p_2)+\mathcal{L}(p_2,p_3)+\mathcal{L}(p_3,p_1))}}{v^T(p_1)\tau_2v(p_2)v^T(p_2)\tau_2 v(p_3) v^T(p_3)\tau_2v(p_1)}.
\end{equation}
When $p_j = \Gamma + \psi_j/\sqrt{N}$, both the numerator and denominator become small, as was the case for the two-point correlation above, since $\mathcal{L}(\Gamma,\Gamma) = 0$ and  $v(\Gamma)^T \tau_2 v(\Gamma) = 0$. However, in the present case the leading term of $\mathcal{O}(N^{3/2})$ of $C_3^{(2,N)}(p)$ vanishes because the leading term of $\mathcal{O}(1/N)$ in $\mathcal{L}(p_j,p_k)$ is real, see~\eqref{eq:L}. It follows that $C_3^{(2,N)}(p) = \mathcal{O}(N)$ only, and therefore the unfolded three-point function vanishes, as observed.

The relative smallness of the three-point correlation function causes the convergence of the winding number distribution to a Gaussian to occur more rapidly than implied by the standard central limit theorem, where the growth of all cumulants is comparable. So the third cumulant of $W$ would be of $\mathcal{O}(N^{1/2})$ with the skewness of the distribution decreasing to zero like $N^{-1/4}$. Here, since $C_3$ is of $\mathcal{O}(N)$, the third cumulant of $W$ is only of $\mathcal{O}(1)$, resulting in the skewness decreasing as $N^{-3/4}$. 

The detailed calculations in Appendix~\ref{secIVB}, starting from \eqref{2WMomentDef}, confirm this observations, and show that the central moments of $W$ become Gaussian in the large $N$ limit with error estimates
\begin{align}	\label{3WMomentsRes}
%\begin{split}
\left\langle (W-\langle W\rangle)^{2\tilde{k}} \right\rangle &= (2\tilde{k}-1)!! N^{\tilde{k}/2} I_2^{\tilde{k}} + \mathcal{O}\left( N^{(\tilde{k}-1)/2} \right),
\\ \label{eq:Wkodd}
\left\langle (W-\langle W\rangle)^{2\tilde{k}+1} \right\rangle &=\mathcal{O}\left( N^{(\tilde{k}-1)/2} \right)
%\end{split}
\end{align}
for even $k=2\tilde{k}$ respectively odd $k=2\tilde{k}+1$ with $\tilde{k}\in\mathbb{N}_0$. We recall that $k!!=k(k-2)(k-4)\cdots$ is the double factorial. Here the integral (variance of $W$ renormalized by $1/\sqrt{N}$)
\begin{align}	\nonumber
I_2 =& \frac{2}{(2\pi)^2} \sum_j \int\limits_0^{2\pi} dt\frac{ \norm{\Gamma_j'(t)}}{\sqrt{|\Delta(\Gamma_{j,1}(t))|^2+|\Delta(\Gamma_{j,2}(t))|^2}} \int\limits_{-\infty}^\infty d\delta p\, f_{2}\left(\Gamma_j(t),\delta p\left[\begin{array}{c} s_1/|\Delta(\Gamma_{j,1}(t))| \\ -s_2/|\Delta(\Gamma_{j,2}(t))| \end{array}\right]\right)
\\	\label{I2}
=& \frac{1}{2\pi^{3/2}} \sum_j \int\limits_0^{2\pi} dt \frac{s_1 s_2 \norm{\Gamma_j'(t)} }{\sqrt{|\Delta(\Gamma_{j,1}(t))|^2+|\Delta(\Gamma_{j,2}(t))|^2}}
\end{align}
emerges, where the summation runs over all curves $\Gamma_j(t)\in[0,2\pi)^2$ such that  $v(\Gamma_{j,1}(t))$ and $v(\Gamma_{j,2}(t))$ are parallel. For this result, we have parametrized
\begin{equation}\label{2.para}
\left[\begin{array}{c} p_1 \\ p_2 \end{array}\right]=\Gamma_j(t)+\frac{\delta p}{\sqrt{N}}\left[\begin{array}{c} s_1/|\Delta(\Gamma_{j,1}(t))| \\ -s_2/|\Delta(\Gamma_{j,2}(t))| \end{array}\right],
\end{equation}
leading to the Jacobian 
\begin{equation}
\lim_{N\to\infty} \sqrt{N} dp_1 dp_2 = \left|\frac{\Gamma'_{j,1}s_2}{|\Delta(\Gamma_{j,2})|}+\frac{\Gamma'_{j,2}s_1}{|\Delta(\Gamma_{j,1})|}\right|dtd\delta p = \frac{2 \norm{\Gamma'_j}}{\sqrt{|\Delta(\Gamma_{j,1})|^2+|\Delta(\Gamma_{j,2})|^2}} dt d\delta p,
\end{equation}
where we have used that $\Gamma'_j$ is proportional to $(s_2|\Delta(\Gamma_{j,2})|,s_1|\Delta(\Gamma_{j,1})|)$, i.e.
\begin{equation}
\Gamma'_j=\frac{\norm{\Gamma'_j}}{\sqrt{|\Delta(\Gamma_{j,1})|^2+|\Delta(\Gamma_{j,2})|^2}}\left[\begin{array}{c} s_2|\Delta(\Gamma_{j,2})| \\ s_1|\Delta(\Gamma_{j,1})| \end{array}\right],
\end{equation}
see comment after~\eqref{Hessian}.

In particular, we find the kurtosis to be
\begin{equation}
\frac{\left\langle (W-\langle W\rangle)^4 \right\rangle}{\left\langle (W-\langle W\rangle)^2 \right\rangle^2}= 3 + o(1).
\end{equation}
The normalized odd moments follow the expression
\begin{equation}
\frac{\left\langle (W-\langle W\rangle)^k \right\rangle}{\left\langle (W-\langle W\rangle)^2 \right\rangle^{k/2}} = \frac{(k-1)\, k!!}{6} \frac{I_3}{I_2^{3/2}} N^{-1/4} + \mathcal{O}\left( N^{-3/4} \right),
\end{equation}
which actually vanishes in the large $N$ limit, since $I_3 = 0$, see Eq.~\eqref{I3}. Therefore, the limiting  winding number distribution is the Gaussian
\begin{equation}
P(W) = \frac{1}{\sqrt{2\pi \sqrt{N}\, I_2}} \exp\left( -\frac{(W-\langle W\rangle)^2}{2 \sqrt{N}\, I_2} \right),
\end{equation}
which is fully determined by $I_2$, see Eq.~\eqref{I2}.

\begin{figure}
\centering
\includegraphics[width=0.8\textwidth]{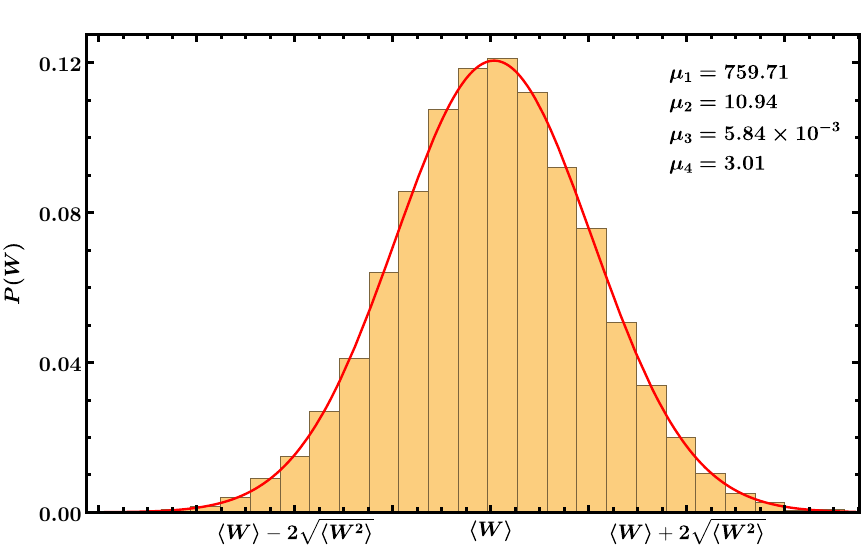}
\caption{Winding number distribution for $10^4$ realizations of a random matrix model~\eqref{model} with $N = 1500$ and parameters $a_1 = 0.92 + 0.82 i,\, a_2 = 0.91 - 0.77 i,\, b_1 = 0.41 - 0.95 i,\, b_2 = -0.84 - 0.70 i$. The left ordinate shows the frequency of each value and the right ordinate the value of the associated probability density function, plotted in red. The inset shows the mean $\mu_1$, the variance $\mu_2$, the skewness $\mu_3$ and the kurtosis $\mu_4$, rounded to two decimals.}
\label{3WNUmDis}
\end{figure}

In Fig. \ref{3WNUmDis} we plot the winding number distribution for the random matrix model
\begin{equation}\label{model}
K(p) = \left(a_1 + a_2 e^{ip} \right) K_1 + \left(b_1 + b_2 e^{ip}\right) K_2
\end{equation}
 with coefficients $a_1, a_2, b_1, b_2 \in \mathbb{C}$ and generic complex $K_1, K_2 \in \mathbb{C}^{N\times N}$. We generated $10^4$ realizations of $K(p)$ with dimension $N = 1500$ and computed the winding number by counting the zeros of $\det K(p)$ as a function of $s = e^{ip}$ inside the unit circle. The numerical result is in good agreement with a Gaussian distribution. 

For the simple choice $v(p) = (\cos p, \sin p)$, the unfolded two-point correlation function is given by
\begin{equation}
f_2(\Gamma_j, \psi) = \frac{\exp \displaystyle \left[- (\psi_1 - \psi_2)^2 \right]
- 1}{ \left(\psi_1-\psi_2 \right)^2}.
\end{equation}
For both solutions $\Gamma_1=(t,t)$ and $\Gamma_2=(t,\pi+t)$ because of $s_1|\Delta(\Gamma_1)|=s_2|\Delta(\Gamma_2)|=1$. Therefore, it is
\begin{equation}
I_2=\frac{1}{2\pi^{3/2}} \sum_j \int\limits_0^{2\pi} dt=\frac{2}{\sqrt{\pi}}
\end{equation}
agreeing with the result of Ref.~\onlinecite{BHWGG2021}. Since the vector $(\cos p, \sin p)$ is already orthogonal to its derivative $ (-\sin p,\cos p)$, the winding number has been centered from the start, i.e. $\langle W\rangle=0$.

\section{Conclusions}
\label{secV}

We studied the winding number statistics in a random matrix model of the chiral unitary class AIII. The starting point of the present work was the generating function of the $k$-point correlation function of the winding number density, which we calculated in a previous work\cite{HKGG2023a} using the method proposed in Ref.~\onlinecite{KieburgGuhr2010a}. 

To facilitate the calculations we centered the winding number distribution with a local transformation of the matrix model. We then showed that, in the large $N$ limit, the correlation function factorizes into blocks over regions in the $k$-dimensional parameter space where the coefficient vectors are parallel. 
This factorization made it possible to compute the $k$th moments of the winding number in leading order of $N$. The leading order terms coincide with the moments of the Gaussian distribution, which proves the Gaussianity of the winding number distribution. We support this finding with numerical results. Furthermore, our results demonstrate that the employed two-matrix model captures universal behaviour in the large $N$ limit, since it depends only on a finite number of functions that can be chosen by varying the coefficient functions of the model.

In Refs. \onlinecite{HKGG2023a, HKGG2023b}, we have also calculated the generating functions for the chiral orthogonal class BDI and the chiral symplectic class CII. In these classes, the underlying random matrix ensemble is real resp. real quaternion, which leads to Pfaffian expressions instead of determinantal ones. The method applied here is likely fruitless in these cases due to the considerably more complicated expressions involved. Choosing a less general random matrix model, such as the trigonometric model $v(p) = \left( \cos p, i\sin p \right)$, while also limiting the calculation to small values of $k$ seems to be a more promising approach.

\begin{acknowledgments}
This work was funded by the German--Israeli Foundation within the project \textit{Statistical Topology of Complex Quantum Systems}, grant number GIF I-1499-303.7/2019 (N.H., O.G. and T.G.). Furthermore, M.K. acknowledges support by the Australian Research Council via Discovery Project grant DP210102887 and the mobility grant funded by the Alexander von Humboldt Foundation, and he is grateful for the hospitality of the Faculty of Physics at Bielefeld University and of the ZiF (the Centre of interdisciplinary Research). N.H. acknowledges support from Chalmers University of Technology via the Area of Advance Nano.
\end{acknowledgments}

\section*{Author Declarations}
\subsection*{Conflict of Interest}
The authors have no conflicts to disclose.

\appendix
\section{Derivations}
\label{secIV}

In Appendix~\ref{secIVA}, we derive the full $k$-point function and perform the large $N$ limit according to definition \eqref{2UnfoldedkPointFunctionDef}, under the assumption that at the expansion point $\Gamma_i$ all vectors $v(\Gamma_{i,n})$ are parallel. In Appendix~\ref{secIVB}, we demonstrate that the correlation function factorizes into lower order correlation functions when this assumption is relaxed. From this decomposition follows the large $N$ estimate for the moments \eqref{3WMomentsRes} and finally the Gaussianity of the winding number distribution.

\subsection{Derivation of the Unfolded \texorpdfstring{$k$-}{}Point Correlation Function}
\label{secIVA}

We need to differentiate the generating function~\eqref{2GenFuncL} once with respect to each source variable $J_n$ at zero. Before doing so, we multiply with
\begin{equation}
1 = \prod_{l=1}^k \frac{v^T(p_n) \tau_2 v(p_n + J_n)}{v^T(p_n) \tau_2 v(p_n + J_n)},
\end{equation}
which yields
\begin{equation}	\label{4GenFunc}
Z^{(2,N)}_{k|k}(p,J) = \frac{\det \left[\displaystyle \frac{v^T(p_n)\tau_2 v(p_n+J_n)}{v^T(p_m)\tau_2 v(p_n+J_n)} \exp[-N\mathcal{L}(p_m,p_n)] \left(\frac{S(p_m,p_n+J_n)}{S(p_m,p_n)} \right)^{N} \right]_{1\leq m,n\leq k}}
{\det \left[\displaystyle \frac{v^T(p_n)\tau_2 v(p_n + J_n)}{v^T(p_m)\tau_2v(p_n+J_n)} \right]_{1\leq m,n\leq k}}.
\end{equation}
This slight modification helps enormously because on the diagonal $m=n$ both original determinants are singular at $J_n=0$ due to the inverse of $v^T(p_n)\tau_2 v(p_n) = 0$, which is remedied by this trick. An expansion in the source variables gives
\begin{equation}
\frac{v^T(p_n)\tau_2 v(p_n+J_n)}{v^T(p_m)\tau_2 v(p_n+J_n)} = \begin{cases}
\displaystyle \frac{1}{S(p_n,p_m)}		\qquad	&v(p_n) \parallel v(p_m),
\\
\displaystyle J_n \frac{\Delta(p_n)}{v^T(p_m) \tau_2 v(p_n)}+ \mathcal{O}\left(J_n^2 \right)		\qquad	&\text{otherwise}.
\end{cases}
\end{equation}
We recall that $\Delta(p_n) = v^T(p_n) \tau_2 v'(p_n)$. When integrating over all $p_j$ and assuming a non-constant modulus of the correlation function $|S(p,q)|$ which is smooth on the square $[0,2\pi)^2$, the vectors $v(p_n)$ and $v(p_m)$ for $n\neq m$ are parallel only on sets of measure zero. Therefore, the generic situation of two parallel vectors $v(p_n)$ and $v(p_m)$ can only occur in the case $n= m$, which allows us to write
\begin{equation}
\frac{v^T(p_n)\tau_2 v(p_n+J_n)}{v^T(p_m)\tau_2 v(p_n+J_n)} =\delta_{nm}+\left[ J_n \frac{\Delta(p_n)}{v^T(p_m) \tau_2 v(p_n)}+ \mathcal{O}\left(J_n^2 \right)\right](1-\delta_{nm}),
\end{equation}
where we use the Kronecker symbol $\delta_{nm}$.

The expression in the numerator benefits from this simple trick, especially when combining it with
\begin{equation}
S(p_n,p_n+J_n)=v^\dagger(p_n)v(p_n+J_n)=1+\mathcal{O}(J_n^2)
\end{equation}
and $\mathcal{L}(p_n,p_n)=0$.
Then, it is for generic $p_n$ and $p_m$ (meaning $v(p_n)$ and $v(p_m)$ are not parallel)
\begin{equation}\label{4OMat}
\begin{split}
&\frac{v^T(p_n)\tau_2 v(p_n+J_n)}{v^T(p_m)\tau_2 v(p_n+J_n)} \exp[-N\mathcal{L}(p_m,p_n)] \left(\frac{S(p_m,p_n+J_n)}{S(p_m,p_n)} \right)^{N}\\
=&\left[1+\mathcal{O}(J_n^2)\right]\delta_{nm}+\biggl[J_n \frac{\Delta(p_n)}{v^T(p_m) \tau_2 v(p_n)}  \exp\left[ -N\mathcal{L} (p_m,p_n ) \right] + \mathcal{O}\left(J_n^2 \right)\biggl](1-\delta_{nm}).
\end{split}
\end{equation}
To shorten the formulas, we define
\begin{equation}	\label{4F1Exp}
F_1 := \det\left[ \delta_{mn} + J_n \frac{\Delta(p_n)}{v^T(p_m) \tau_2 v(p_n)}  \exp\left[-N\mathcal{L}(p_m,p_n) \right] (1-\delta_{nm}) \right]_{1\leq m,n\leq k}
\end{equation}
for the numerator of the generating function, which includes only terms up to order one in each source variable.

Similarly, we can deal with the denominator, using the Cauchy determinant identity, see, e.g. Ref.~\onlinecite{KieburgGuhr2010a}. Recalling $v(p)=(a(p),b(p))$ and defining the abbreviation $\kappa(p)=a(p)/b(p)$, we compute
\begin{align}\label{Cauchy.det}
\det^{-1} \left[\displaystyle \frac{v^T(p_n)\tau_2 v(p_n + J_n)}{v^T(p_m)\tau_2v(p_n+J_n)} \right]_{1\leq m,n\leq k}=&\det^{-1} \left[ \frac{\kappa(p_n+J_n) - \kappa(p_n)}{\kappa(p_n+J_n) - \kappa(p_m)} \right]_{1\leq m,n\leq k}
\\	\nonumber
=&\prod_{m<n} \left( \frac{\kappa(p_m+J_m) - \kappa(p_n)}{\kappa(p_m+J_m) - \kappa(p_n+J_n)} \frac{\kappa(p_n+J_n) - \kappa(p_m)}{\kappa(p_n) - \kappa(p_m)} \right)
\\	\nonumber
=& \prod_{m<n} \left( 1 - J_m J_n \frac{\kappa'(p_m) \kappa'(p_n)}{[\kappa(p_m) - \kappa(p_n)]^2} + \mathcal{O}\left(J_m^2 +J_n^2\right) \right)
\\	\nonumber
=& \prod_{m<n} \left( 1 - J_m J_n \frac{\Delta(p_m) \Delta(p_n)}{[ v^T(p_m) \tau_2 v(p_n)]^2} + \mathcal{O}\left(J_m^2 +J_n^2\right)\right).
\end{align}
Anew, we drop all higher order terms in each $J_n$ and define
\begin{equation}\label{F2.def}
F_2:=\prod_{m<n} \left( 1 - J_m J_n \frac{\Delta(p_m) \Delta(p_n)}{[ v^T(p_m) \tau_2 v(p_n)]^2} \right)
\end{equation}
to shorten the expressions.

Hence, the correlation function can be obtained from the derivative of the product $F_1 F_2$
\begin{equation}\label{Ck2N}
C_k^{(2,N)}(p) = \left(\prod_{l=1}^k \frac{\partial}{\partial J_l}\right) Z^{(2,N)}_{k|k}(p, J) \Bigg|_{J=0} = \left(\prod_{l=1}^k  \frac{\partial}{\partial J_l} \right) F_1 F_2 \Bigg|_{J=0}
\end{equation}
for which the product rule has to be invoked.

Any derivative of the numerator $F_1$ can be worked out directly with the help of the multilinearity of the determinant and the Kronecker symbols. For any set $\{l_1, \ldots, l_m \} \subseteq \{1,\ldots, k\}$ of $m\leq k$ pairwise distinct elements it is
\begin{equation}\label{F1.der}
\begin{split}
\frac{\partial^m}{\partial J_{l_1} \cdots \partial J_{l_m}} F_1 \Bigg|_{J=0} 
%&= \frac{\partial^m}{\partial J_{l_1} \cdots \partial J_{l_m}} \sum_{\sigma \in \mathbb{S}_k} \sgn\, \sigma \prod_{n=1}^k \left[\delta_{\sigma(n) n} + J_n O_{\sigma(n)n}(p_{\sigma(n)}, p_n) \right] \Bigg|_{J=0}\\&= \sum_{\sigma \in \mathbb{S}_k} \sgn\, \sigma \prod_{n \in \{1,\ldots,k\} \setminus \{l_1,\ldots, l_m\}} \delta_{\sigma(n) n} \prod_{n \in \{l_1,\ldots, l_m\}} O_{\sigma(n)n}(p_{\sigma(n)}, p_n) \\&= \sum_{\sigma \in \mathbb{S}_{k-m}} \sgn\, \sigma \prod_{n \in \{l_1,\ldots, l_m\}} O_{\sigma(n)n}(p_{\sigma(n)}, p_n) 
= \det\left[ \frac{\Delta(p_b)}{v^T(p_a) \tau_2 v(p_b)}  \exp\left[-N\mathcal{L}(p_a,p_b) \right] (1-\delta_{ab})\right]_{a,b \in \{l_1, \ldots, l_m \}}.
\end{split}
\end{equation}
Due to the factor $(1-\delta_{ab})$, we have for $m=2$ and $m=3$
\begin{equation}\label{O.m2.3}
\begin{split}
\frac{\partial^2}{\partial J_{l_1} \partial J_{l_2}} F_1 \Bigg|_{J=0} =&\frac{\Delta(p_{l_1})\Delta(p_{l_2})}{[v^T(p_{l_1}) \tau_2 v(p_{l_2})]^2} e^{-N[\mathcal{L}(p_{l_1},p_{l_2})+\mathcal{L}(p_{l_2},p_{l_1})] },
\\
\frac{\partial^3}{\partial J_{l_1} \partial J_{l_2}\partial J_{l_3}} F_1 \Bigg|_{J=0} =&\frac{\Delta(p_{l_1})\Delta(p_{l_2})\Delta(p_{l_3})}{v^T(p_{l_1}) \tau_2 v(p_{l_2}) v^T(p_{l_2}) \tau_2 v(p_{l_3}) v^T(p_{l_3}) \tau_2 v(p_{l_1})}
\\
&\hspace{-1.5cm} \times \left( e^{-N[\mathcal{L}(p_{l_1},p_{l_2})+\mathcal{L}(p_{l_2},p_{l_3})+\mathcal{L}(p_{l_3},p_{l_1})]}-e^{-N[\mathcal{L}(p_{l_1},p_{l_3})+\mathcal{L}(p_{l_3},p_{l_2})+\mathcal{L}(p_{l_2},p_{l_1})]}\right),
\end{split}
\end{equation}
which represent a drastic simplification compared to the starting equation.

 We note that each factor in $F_2$ depends on pairs of $J_n$. It follows that only derivatives of $F_2$ that are of even order contribute. Each of the pairs yields a factor $\Delta(p_m) \Delta(p_n)/[v^T(p_m) \tau_2 v(p_n)]^2$.
We construct the correlation function as a sum over the number of pairs $l$, ranging from $0$ to $\lfloor k/2 \rfloor$. To account for all possible combinations of derivatives appearing in the product rule, we sum over all permutations of the indices.
However, this results in an overcounting due to the permutation invariance of the individual terms, which we need to amend by multiplying with combinatorial factors. The product of the pairs above  are invariant under transposition in each pair yielding a factor $1/2^l$. The permutations of the pairs yield a factor $1/l!$ and the permutations of the remaining variables lead to a factor $1/(k-2l)!$. Once this combinatorial overcounting is taken into account, we obtain the $k$-point correlation function~\eqref{4CorrFuncRes}
\begin{align}
C_k^{(2,N)}(p) = \sum_{l=0}^{\lfloor k/2 \rfloor} \frac{(-1)^l}{2^l\, l!\, (k-2l)!} &\sum_{\sigma \in \mathbb{S}_k} \left( \prod_{m=1}^l \frac{\Delta \left(p_{\sigma(2m-1)} \right) \Delta \left(p_{\sigma(2m)} \right) }{ \left[ v^T \left(p_{\sigma(2m-1)} \right) \tau_2 v \left(p_{\sigma(2m)} \right) \right]^2} \right) 
\\	\nonumber
\times &\det\left[ \frac{\Delta(p_{\sigma(b)})}{v^T(p_{\sigma(a)}) \tau_2 v(p_{\sigma(b)})}  e^{-N\mathcal{L}\left( p_{\sigma(a)},p_{\sigma(b)} \right)} (1-\delta_{ab})\right]_{a,b = 2l+1,\ldots, k},
\end{align}
which is valid for any $k$.

We now turn to the large $N$ limit, using definition~\eqref{2UnfoldedkPointFunctionDef} of the unfolded $k$-point function 
\begin{equation}
f_k(\Gamma, \psi) = \lim_{N \to \infty} N^{-k/2} C_k^{(2,N)} \left(\Gamma +\frac{ \psi}{\sqrt{N}} \right)
\end{equation}
and assume that we expand around the points $\Gamma_j(t)=(\Gamma_{j,1}(t),\ldots,\Gamma_{j,k}(t)) \in [0,2\pi)^k$ where all vectors $v(\Gamma_{j,n}(t) )$ are parallel for each fixed $t\in[0,2\pi)$. The other cases are covered in Appendix~\ref{secIVB}. The expansion is given by $p_a = \Gamma_{j,a}(t) + \psi_a/\sqrt{N}$ where the vector $\psi=( \psi_1,\ldots, \psi_k)$ parametrizes the $(k-1)$-dimensional subspace orthogonal to $\Gamma_j'(t)=(\Gamma_{j,1}'(t),\ldots,\Gamma_{j,k}'(t))$.

The parallelism of the vectors gives $v^T(\Gamma_{j,m}) \tau_2 v(\Gamma_{j,n})=0$ so that
\begin{equation}
v^T(p_m) \tau_2 v(p_n)=\frac{1}{\sqrt{N}}\left( v^T(\Gamma_{j,m}) \tau_2 v'(\Gamma_{j,n})\psi_n-v^T(\Gamma_{j,n}) \tau_2 v'(\Gamma_{j,m})\psi_m \right)
+ \mathcal{O}\left( \frac{1}{N} \right).
\end{equation}
With relation \eqref{proportionality} we can rewrite this expansion as follows
\begin{equation}
v^T(p_m) \tau_2 v(p_n)=\frac{1}{\sqrt{N}}\left( S(\Gamma_{j,m},\Gamma_{j,n})\Delta(\Gamma_{j,m})\psi_m-S(\Gamma_{j,n},\Gamma_{j,m})\Delta(\Gamma_{j,n})\psi_n \right)
+ \mathcal{O}\left( \frac{1}{N} \right).
\end{equation}
We recall that $S(\Gamma_{j,m},\Gamma_{j,n})=S^*(\Gamma_{j,n},\Gamma_{j,m})$ is only a complex phase.
Plugging this expansion into the factors in the $k$-point correlation function~\eqref{4CorrFuncRes} we obtain
\begin{equation}
\label{41stFactorExpansion}
\frac{\Delta(p_m) \Delta(p_n)}{\left[ v^T(p_m) \tau_2 v(p_n) \right]^2} 
= N \frac{\Delta(\Gamma_{j,m})\Delta(\Gamma_{j,n})}{ \left( S(\Gamma_{j,m},\Gamma_{j,n})\Delta(\Gamma_{j,m})\psi_m-S(\Gamma_{j,n},\Gamma_{j,m})\Delta(\Gamma_{j,n})\psi_n \right)^2}\left[ 1+ \mathcal{O}\left(\frac{|\psi_m-\psi_n|}{\sqrt{N}}\right) \right],
\end{equation}
where we used the parallelism assumption for the expansion of the denominator. We proceed by exploiting~\eqref{Delta.rel}, where we replace $dp/dq$ by $\Gamma'_{j,n}/\Gamma'_{j,m}$ to arrive at 
\begin{equation}	\label{41stFactorExpansion.b}
\begin{split}
\frac{\Delta(p_m) \Delta(p_n)}{[v^T(p_m) \tau_2 v(p_n)]^2} 
=& N \frac{\Gamma'_{j,n}\Gamma'_{j,m}}{ \left( \Gamma'_{j,n}\psi_m-\Gamma'_{j,m}\psi_n \right)^2} \left[ 1+ \mathcal{O}\left(\frac{|\psi_m-\psi_n|}{\sqrt{N}}\right) \right]
\\
=&N \frac{|\Delta(\Gamma_{j,m})|\,|\Delta(\Gamma_{j,n})|}{ \left(s_{j,m} |\Delta(\Gamma_{j,m})|\psi_m-s_{j,n}|\Delta(\Gamma_{j,n})|\psi_n \right)^2} \left[ 1+ \mathcal{O}\left(\frac{|\psi_m-\psi_n|}{\sqrt{N}}\right) \right],
\end{split}
\end{equation}
which expresses everything in terms of the derivative $\Gamma_j'$ or, alternatively, in terms of $|\Delta(p)|=\sqrt{\partial_1\partial_2 S(p,p)}$. We have employed the abbreviation $s_{j,m} = \text{sign }\Gamma'_{j,m}$ as well as that $(\Gamma'_{j,n},-\Gamma'_{j,m})$ must be proportional to $(s_{j,m}|\Delta(\Gamma_{j,m})|,-s_{j,n}|\Delta(\Gamma_{j,n})|)$, see comment after~\eqref{sec.eig}.

Next, we expand the elements of the determinant~\eqref{4OMat}. Using~\eqref{2LExpApprox}, we introduce the quantity
\begin{align}	\label{4OMatExpansion}
O(p_m,p_n)=&\frac{\Delta(p_n)}{v^T(p_m) \tau_2 v(p_n)}  \exp\left[-N\mathcal{L}(p_m,p_n) \right]
\\	\nonumber
=& \sqrt{N} \frac{1}{S(\Gamma_{j,n}, \Gamma_{j,m})} \exp\left[  -\frac{|\Delta(\Gamma_{j,m})|^2+|\Delta(\Gamma_{j,n})|^2}{2} \frac{\left( \Gamma'_{j,n}\psi_m-\Gamma'_{j,m}\psi_n \right)^2}{ (\Gamma'_{j,m})^2+ ( \Gamma'_{j,n})^2}\right] 
\\	\nonumber
&\times \frac{\Gamma'_{j,n}}{\Gamma'_{j,n}\psi_m-\Gamma'_{j,m}\psi_n} \left[ 1+ \mathcal{O}\left(\frac{(|\psi_m|^2+|\psi_n|^2)|\psi_m-\psi_n|}{\sqrt{N}}\right) \right]
\\	\nonumber
=& \sqrt{N} \frac{1}{S(\Gamma_{j,n}, \Gamma_{j,m})} \frac{s_{j,m} |\Delta(\Gamma_{j,m})|}{s_{j,m} |\Delta(\Gamma_{j,m})|\psi_{j,m}-s_{j,n}|\Delta(\Gamma_{j,n})|\psi_n}
\\	\nonumber
&\times \exp\left[ -\frac{\left(s_{j,m} |\Delta(\Gamma_{j,m})|\psi_m- s_{j,n}|\Delta(\Gamma_{j,n})|\psi_n \right)^2}{2}\right] \left[ 1 + \mathcal{O}\left(\frac{(|\psi_m|^2+|\psi_n|^2)|\psi_m-\psi_n|}{\sqrt{N}}\right) \right].
\end{align}
The complex phase $S(\Gamma_{j,n}, \Gamma_{j,m})$ actually drops out once we plug~\eqref{4OMatExpansion} into the determinant~\eqref{F1.der}. To see this, we exploit the parallelism of the vectors $v(\Gamma_{j,m})$. This leads to the transitivity relation
\begin{equation}
S(\Gamma_{j,m},\Gamma_{j,n})=v^\dagger(\Gamma_{j,m})v(\Gamma_{j,n})=S(\Gamma_{j,l},\Gamma_{j,n})v^\dagger(\Gamma_{j,m})v(\Gamma_{j,l})=S(\Gamma_{j,m},\Gamma_{j,l})S(\Gamma_{j,l},\Gamma_{j,n}),
\end{equation}
because of the relation $v=(w^\dagger v) w$ whenever two normalized vectors $v$ and $w$ are parallel.
As a determinant can be understood as a sum (up to signs) over products of cycles of matrix entries we have
\begin{align}\label{F1.der.b}
&\frac{\partial^m}{\partial J_{l_1} \cdots \partial J_{l_m}} F_1 \Bigg|_{J=0}
\\	\nonumber
=&N^{m/2} \prod_{i=1}^m s_{j,l_i}|\Delta(\Gamma_{j,l_i})|\, \det\left[ \frac{\displaystyle  \exp\left[ -\frac{1}{2} \left(s_a|\Delta(\Gamma_{j,a})|\psi_{a}-s_{b}|\Delta(\Gamma_{j,b})|\psi_b \right)^2 \right]}{s_a|\Delta(\Gamma_{j,a})|\psi_a-s_{b}|\Delta(\Gamma_{j,b})|\psi_b} (1-\delta_{ab})\right]_{a,b \in \{l_1, \ldots, l_m \}}
\\	\nonumber
&\times\left[ 1+ \sum_{a<b}\mathcal{O}\left(\frac{(|\psi_{l_a}^2+|\psi_{l_b}|^2)|\psi_{l_a}-\psi_{l_b}|}{\sqrt{N}}\right) \right].
\end{align}
Evidently, the determinant vanishes when $m$ is odd as we then take a determinant of an odd-dimensional antisymmetric matrix. Combining the intermediate results~\eqref{Ck2N},~\eqref{41stFactorExpansion.b} and~\eqref{F1.der.b} we arrive at~\eqref{3UnfoldedCorrRes}.

\subsection{Decomposition of the Correlation Function and Statistical Moments}
\label{secIVB}

For the moment we will drop the assumption that the vectors $v(p_l)$ are parallel to see that actually a clustering into pairs of points will be the leading contribution. The vector $p=(p_1,\ldots,p_k) \in [0,2\pi)^k$ shall be arbitrary, which does not exclude the possibility of parallel vectors. In the following, we expand the generating function $Z^{(2,N)}_{k|k}\left( p, J/\sqrt{N} \right)$
in leading order of $N$. We introduced a rescaling of the source variables, which is convenient when doing the expansions, and is compensated by a prefactor when evaluating the correlation function
\begin{equation}
C^{(2,N)}_k \left( p \right) = N^{k/2} \left( \prod_{l=1}^k \frac{\partial}{ \partial J_l} \right)  Z^{(2,N)}_{k|k} \left( p, J/\sqrt{N} \right) \Bigg|_{J=0}.
\end{equation}
To deal with the following analysis, we call two vectors not close to being parallel when $1-|S(p_a,p_b)| \gg 1/N$, otherwise we call them close to being parallel. We recall that $0 \leq |S(p_a,p_b)| \leq 1$.

The following procedure works along the same lines as in the last subsection. Everything depends on whether $v(p_a)$ and $v(p_b)$ are close to being parallel or not because of the decomposition
\begin{equation}
v(p_m)=S(p_n,p_m)v(p_n)+\sqrt{1 - |S(p_n,p_m)|^2} \frac{v^T(p_n)\tau_2v(p_m)}{|v^T(p_n) \tau_2 v(p_m)|} \tau_2 v^*(p_n).
\end{equation}
We have the following expansion of the matrix in the determinant in the denominator of~\eqref{4GenFunc}
\begin{equation}\label{1st.exp}
\begin{split}
&\frac{v^T(p_n) \tau_2 v(p_n+ J_n/\sqrt{N})}{v^T(p_m) \tau_2 v(p_n + J_n/\sqrt{N})}
\\
=&\delta_{nm}+(1-\delta_{nm})\left[J_n \frac{\Delta(p_n) }{\sqrt{N}\sqrt{1-|S(p_n,p_m)|^2}}\frac{|v^T(p_m)\tau_2v(p_n)|}{v^T(p_m)\tau_2v(p_n)}+\mathcal{O}\left(J_n^2\right) \right].
\end{split}
\end{equation}
The off-diagonal terms evidently vanish when $v(p_m)$ and $v(p_n)$ are not close to being parallel while it is of order one or larger when $1-|S(p_n,p_m)| \gg 1/N$. Similarly we can deal with the matrix entries of the determinant in the numerator of~\eqref{4GenFunc} which is
\begin{align}\label{2nd.exp}
&\frac{v^T(p_n)\tau_2 v(p_n+J_n)}{v^T(p_m)\tau_2 v(p_n+J_n)} \exp[-N\mathcal{L}(p_m,p_n)] \left(\frac{S(p_m,p_n+J_n)}{S(p_m,p_n)} \right)^{N}=[1+\mathcal{O}\left(J_n^2\right)]\delta_{nm}
\\	\nonumber
&+(1-\delta_{nm})\left[J_n \frac{\Delta(p_n) }{\sqrt{N}\sqrt{1-|S(p_n,p_m)|^2}}\frac{|v^T(p_m)\tau_2v(p_n)|}{v^T(p_m)\tau_2v(p_n)}\left(\frac{S(p_m,p_n)}{|S(p_m,p_n)|}\right)^N|S(p_m,p_n)|^N+\mathcal{O}\left(J_n^2\right) \right].
\end{align}
This time, the off-diagonals vanish even exponentially when $1-|S(p_n,p_m)|\gg 1/N$ due to the term $|S(p_n, p_m)|^N\to \exp[-N (1-|S(p_n,p_m)|) ]\to0$ when $N \to \infty$. When  $1-|S(p_n,p_m)|=\mathcal{O}(1/N)$ those terms are of the same order as in~\eqref{1st.exp}.

Making use of this analysis, we bunch the variables $\{ p_1, \ldots, p_k \}$ together into blocks
\begin{equation}
B_l = \{ p_{j_{1,l}}, \ldots, p_{j_{k_l,l}} \},
\end{equation}
where  all vectors $v(p_{j_{1,l}}), \ldots, v(p_{j_{k_l,l}})$ are close to being parallel. We assume that there are $L \leq k$ blocks and that the $l$th block has length $k_l$ so that $\sum_{l=1}^L k_l = k$. For finite $N$, the property "close to being parallel" is not an equivalence relation as transitivity is not satisfied. However, in the limit of large $N$ the condition $1 - |S(p_n, p_m)| \gg 1/N$ enforces parallelism, which is indeed an equivalence relation. Therefore the blocks will be equivalence classes in this limit and it is $B_{l}\cap B_{l'}=\emptyset$ whenever $l\neq l'$ and $\bigcup_{l=1}^L B_l=\{p_1,\ldots,p_k\}$. The analysis above tells us that the generating function factorizes in leading order
\begin{equation}
\begin{split}
Z^{(2,N)}_{k|k}\left( p,\frac{ J}{\sqrt{N} }\right)= \prod_{l=1}^L &Z^{(2,N)}_{k_l|k_l} \left(p_{j_{1,l}},\ldots,p_{j_{k_l,l}},\frac{ J}{\sqrt{N} }\right) 
\\
\times &\left[1 + \mathcal{O}\left(\sum_{j=1}^k J_j^2+\sum_{l<l'}\frac{1}{\sqrt{N \left(1-|S(p_{j_{1,l}},p_{j_{1,l'}})| \right)}}\right)\right].
\end{split}
\end{equation}
For the error, it is more than enough to use $1-|S(p_{j_{1,l}},p_{j_{1,l'}})|$ since this difference is of the same order as  for any other pair $p_a\in B_l$ and $p_b\in B_{l'}$. Therefore the corresponding $k$-point correlation function also factorizes
\begin{equation}	\label{4CorrFuncDecomp}
C^{(2,N)}_k \left(p \right) = \prod_{l=1}^L C^{(2,N)}_{k_l} \left( p_{j_{1,l}},\ldots,p_{j_{k_l,l}} \right) \left[1 + \mathcal{O}\left(\sum_{l<l'}\frac{1}{\sqrt{N \left( 1-|S(p_{j_{1,l}},p_{j_{1,l'}})| \right)}}\right)\right].
\end{equation}
 Combining this with the insights of the previous subjection, each block must follow along a solution $\Gamma_l \in [0,2\pi)^{k_l}$ where all $v(\Gamma_{l,1}),\ldots,v(\Gamma_{l,k_l})$ are parallel. To account for deviations from being exactly parallel and of close to being parallel, we set
\begin{equation}\label{parametrisation}
(p_{j_{1,l}},\ldots,p_{j_{k_l,l}})=\Gamma_l(t_l)+\frac{1}{\sqrt{N}} (\psi_{j_{1,l}},\ldots, \psi_{j_{k_l,l}})
\end{equation}
with $\sum_{a=1}^{k_l} \Gamma'_{l,a}(t_l) \psi_{j_{a,l}} = 0$ and $t_l \in [0,2\pi)$. The latter condition only reflects that we need to parametrize the $(k_l-1)$-dimensional vector space orthogonal to the direction of the solution $\Gamma_l(t_l)$.
One can readily check that close to being parallel is given when $\psi_{a}$ is maximal of order one. To see this, we recall that $v^\dagger(\Gamma_{l,a})v'(\Gamma_{l,b})=0$ for all $a,b=1,\ldots,k_l$. Then, the limit becomes
\begin{equation}	\label{local.4CorrFuncDecomp}
\lim_{N\to\infty} N^{-k/2}C^{(2,N)}_k \left(p \right) = \prod_{l=1}^L f_{k_l} \left(\Gamma_l, \psi_{j_{1,l}},\ldots,\psi_{j_{k_l,l}} \right) .
\end{equation}
We can combine the last $k_l$ arguments of $f_{k_l}$ as a single $k_l$ dimensional vector.
This is the corresponding result when not all $v(p_a)$ are not close to being parallel. In the extreme case, when all $v(p_a)$ are not close to being parallel it is $\lim_{N\to\infty} N^{-k/2}C^{(2,N)}_k \left(p \right)=\prod_{j=1}^k f_1(p_j,0)=0$ because of $f_1(p_j,0) = 0$. The latter also implies that each block $B_l$ must contain at least two elements.

We want to exploit this factorisation by splitting the original integration of $p=(p_1,\ldots,p_k)\in [0,2\pi)^k$ into integrations where $v(p_a)$ are close to being parallel or not.  For this purpose we introduce the characteristic functions $\chi(v(p_m) \parallel v(p_n))$, which is unity when $v(p_m)$ and $v(p_n)$ are close to being parallel and zero otherwise, $\chi(v(p_m) \nparallel v(p_n))=1-\chi(v(p_m) \parallel v(p_n))$. This means for the centered winding number
\begin{equation}
\left\langle (W-\langle W\rangle)^k \right\rangle = \int\limits_{[0,2\pi)^k} \frac{d^kp}{(2\pi i)^k} C^{(2,N)}_k (p)\prod_{n<m} \left[\chi(v(p_m) \parallel v(p_n))+\chi(v(p_m) \nparallel v(p_n))\right].
\end{equation}
When expanding the product we may exploit the permutation invariance of $C^{(2,N)}_k (p)$ in its arguments so that
\begin{align}
\left\langle (W-\langle W\rangle)^k \right\rangle =& \sum_{L=1}^k \sum_{\substack{1\leq k_1\leq \ldots \leq k_L \leq k, \\ \sum_{l=1}^Lk_l = k}} 
\frac{k!}{\prod_{l=1}^L k_l! \prod_{l = 1}^k M_l!}  \int\limits_{[0,2\pi)^k}\frac{d^kp}{(2\pi i)^k} C^{(2,N)}_k (p)
\\	\nonumber
&\times \left[ \prod_{l = 1}^L \prod_{1\leq m < n \leq k_l} \chi (v(p_{m,l}) \parallel v(p_{n,l})) \right] 
\left[ \prod_{1 \leq l < l' \leq L} \prod_{\substack{1\leq m \leq k_{l}, \\ 1\leq n \leq k_{l'}}}
\chi(v(p_{m,l}) \nparallel v(p_{n,l'}) ) \right],
\end{align}
where $p_{n,l} = p_{\sum_{l'=1}^{l-1}k_{l'}+n}$ is the $n$th variable in the $l$th block and $M_l$ is the number of blocks with length $l$. The combinatorial factor $k!/\prod_{l=1}^L k_l!$ is the coefficient of the multinomial expansion and $1/\prod_{l = 1}^k M_l!$ corrects the overcounting due to the permutation invariance within each block.

Next, we apply the parametrization~\eqref{parametrisation} for each block $\{p_{1,l} ,\ldots, p_{k_l,l} \}$. The corresponding Jacobian is equal to
\begin{equation}
d^kp=\prod_{l=1}^LN^{-(k_l-1)/2}\frac{\norm{\Gamma'_{l}(t_l)}^2}{|\Gamma'_{l,k_l}(t_l)|} dt_l d\psi_{1,l} \cdots d\psi_{k_l-1,l}
\qquad {\rm with}\ \norm{\Gamma'_{l}}^2=\sum_{a=1}^{k_l}[\Gamma'_{l,a}(t_l)]^2,
\end{equation}
where we assumed $\sum_{a=1}^{k_l}\Gamma'_{l,a}(t_l)\psi_{j_{k,l}}=0$. We denote the set of solutions of curves in $[0,2\pi)^k$, for which all  $v(\Gamma_{l,1}),\ldots,v(\Gamma_{l,k_l})$ are parallel, by $\Sigma_{k_l} \in [0,2\pi)^{k_l}$. Then it is
\begin{align}	\nonumber
\left\langle (W-\langle W\rangle)^k \right\rangle =& \sum_{L=1}^k \sum_{\substack{2\leq k_1\leq \ldots \leq k_L \leq k, \\ \sum_{l=1}^Lk_l = k}} 
N^{k/2}\frac{k!}{\prod_{l=1}^L k_l! \prod_{l = 1}^k M_l!} \prod_{l=1}^L \sum\limits_{\Gamma_l\in\Sigma_{k_l}}\int\limits_0^{2\pi} \frac{dt_l}{(2\pi i)^{k_l}}\int\limits_{\mathbb{R}^{k_l-1}} \frac{d\psi_{1,l}\cdots d\psi_{k_l-1,l}}{N^{(k_l-1)/2}}
\\	\label{A2WMoments}
&\times \frac{\norm{\Gamma'_{l}(t_l)}^2}{|\Gamma'_{l,k_l}(t_l)|}f_{k_l} (\Gamma_l(t_l),\psi_{1,l} ,\ldots, \psi_{k_l,l}) [1+o(1)].
\end{align}
We already employed that $f_1(p_j,0)=0$ which is reflected in $k_j\geq2$.

Close inspection of this expression shows that the term for the  fixed choice of summing indices $k_1,\ldots,k_L$ scales like $N^{L/2}$. Thus, the larger $L$ is the bigger the contribution. Since $k_a\geq2$ for all $a=1,\ldots,L$ the largest $L$ is given by $\lfloor k/2\rfloor$ with the floor function $\lfloor .\rfloor$. For even $k=2\tilde{k}$, this means $L=\tilde{k}$ and $k_1=\ldots=k_{\tilde{k}}=2$ yielding
\begin{equation}
\begin{split}
\lim_{N\to\infty}N^{-\tilde{k}/2}\left\langle (W-\langle W\rangle)^{2\tilde{k}} \right\rangle =& \frac{(2\tilde{k})!}{2^{\tilde{k}}\,\tilde{k}!} I_2^{\tilde{k}}.
\end{split}
\end{equation}
with
\begin{equation}	\label{I22}
I_2=\sum\limits_{\Gamma\in\Sigma_{2}}\int\limits_0^{2\pi} \frac{dt}{(2\pi i)^{2}}\int\limits_{-\infty}^\infty d\psi\frac{\norm{\Gamma'(t)}^2}{|\Gamma'_2(t)|}f_{2} \left(\Gamma(t),\psi,-\frac{\Gamma'_1(t)}{\Gamma'_2(t)}\psi\right).
\end{equation}
Once we substitute the variables~\eqref{2.para} we arrive at~\eqref{I2} where we combine the two last arguments of $f_2$ into a single vector. For odd $k=2\tilde{k}+1$, it is $L=\tilde{k}$ and $k_1=\ldots=k_{\tilde{k}-1}=2$ and $k_{\tilde{k}}=3$ which results in the normalized centered third moment
\begin{equation}
\begin{split}
\lim_{N\to\infty}N^{1/4}\frac{\left\langle (W-\langle W\rangle)^{2\tilde{k}+1} \right\rangle}{\left\langle (W-\langle W\rangle)^{2} \right\rangle^{(2\tilde{k}+1)/2}} =& \frac{1}{3}\frac{(2\tilde{k}+1)!}{2^{\tilde{k}}\,(\tilde{k}-1)!} \frac{I_3}{I_2^{3/2}}
\end{split}
\end{equation}
with 
\begin{equation}\label{I3}
I_3=\sum\limits_{\Gamma\in\Sigma_{3}}\int\limits_0^{2\pi} \frac{dt}{(2\pi i)^{3}}\int\limits_{\mathbb{R}^{2}} d\psi_{1}d\psi_{2} \frac{\norm{\Gamma'(t)}^2}{|\Gamma'_3(t)|}f_{3} \left(\Gamma(t),\psi_1,\psi_2,-\frac{\Gamma'_1(t)\psi_1+\Gamma'_2(t)\psi_2}{\Gamma'_3(t)}\right)=0
\end{equation}
because of $f_3(\Gamma(t),\psi)=0$, see the remark below~\eqref{3UnfoldedCorrRes}. The term $I_3$ is essentially the third centered moment of the winding number, and we need to go one order higher in the expansion shown in Appendix~\ref{secIVA} to get something non-vanishing. 

The integrands in \eqref{A2WMoments}, \eqref{I22} and \eqref{I3} appear to have an asymmetric dependence on the components of $\Gamma(t)$, as only one component appears in the denominator. This is merely apparent as the solutions are invariant under permutation of the components. We recall that parallelism is an equivalence relation which implies the permutation invariance.

\bibliography{lit}

\end{document}